# Ionization of Ammonia Nanoices With Adsorbed Methanol Molecules

Michal Fárník[*,¶], Andriy Pysanenko[†], Kamila Moriová[†,§], Lorenz Ballauf[‡], Paul Scheier[‡], Jan Chalabala[¶], and Petr Slavíček[*,¶]

J. Heyrovský Institute of Physical Chemistry, The Czech Academy of Sciences, Dolejškova 3, 182 23 Prague, Czech Republic, Institut für Ionenphysik und Angewandte Physik, Universität Innsbruck, Technikerstr. 25, A-6020 Innsbruck, Austria, and Department of Physical Chemistry, University of Chemistry and Technology, Technická 5, 166 28 Prague, Czech Republic

E-mail: michal.farnik@jh-inst.cas.cz; petr.slavicek@vscht.cz

Phone: +0420 2 6605 3206. Fax: +420 2 8658 2307

## Abstract

Large ammonia clusters represent a model system of ices which are omnipresent throughout the space. The interaction of ammonia ices with other hydrogen-boding molecules such as methanol or water and their behavior upon an ionization are thus relevant in the astrochemical context. In this study, ammonia clusters $(NH_3)_N$ with the mean size $\bar{N} \approx 230$ were prepared in molecular beams and passed through a pickup cell in which methanol molecules were adsorbed. At the highest exploited pickup pressures, the average composition of $(NH_3)_N(CH_3OH)_M$ clusters was estimated to be $N:M \approx 210:10$. On the other hand, the electron ionization of these clusters yielded about 75% of methanol-containing fragments $(NH_3)_n(CH_3OH)_m H^+$ compared to 25% contribution of pure ammonia $(NH_3)_n H^+$ ions. Based on this substantial disproportion, we propose the following ionization mechanism: The prevailing ammonia is ionized in most cases, resulting in $NH_4^+$ core solvated most likely with four ammonia molecules, yielding the well-known "magic number" structure $(NH_3)_4NH_4^+$. The methanol molecules exhibit strong propensity for sticking to the fragment ion. We have also considered mechanisms of intracluster reactions. In most cases, proton transfer between ammonia units take place. The theoretical calculations suggested the proton transfer either from the methyl group or from the hydroxyl group of the ionized methanol molecule to ammonia to be the energetically open channels. However, the experiments with selectively deuterated methanols did not show any evidence for the $D^+$ transfer from the $CD_3$ group. The proton transfer from the hydroxyl group could not be excluded entirely nor confirmed unambiguously by the experiment.

# Introduction

We investigate the electron impact induced reactions in mixed ammonia-methanol clusters. Molecular clusters are often considered as a bridge between isolated molecule and condensed phase[1] - in the present case, the large ammonia clusters represent a proxy to ammonia ices with methanol molecules on its surface. Such systems can play a role in astrochemistry.

Complexes of methanol and ammonia represent a non-trivial example of hydrogen bonded (HB) systems with various potential reaction pathways upon ionization. The generic feature for many of the ionized hydrogen bonded systems is a proton transfer reaction (PT), taking place from the hydrogen bond donor to the hydrogen bond acceptor unit:

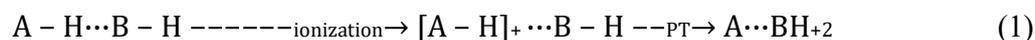

$$A - H \cdots B - H \xrightarrow{\text{ionization}} [A - H]_+ \cdots B - H \xrightarrow{\text{PT}} A \cdots BH_{+2} \qquad (1)$$

where A and B can be generally different electronegative atoms or groups.[2–5] The proton transfer reaction was most widely studied in various forms of water, i.e. in water clusters,[6–17] liquid water[18] or ices.[19,20] The proton transfer in water belongs among the fastest chemical processes - the estimates for the $H_2O^+$ lifetime are below 100 fs.[21,22]

Water forms a very compact hydrogen bond network and the proton transfer process might be less efficient in other hydrogen bonded systems. Indeed, the HB in ammonia aggregates is much less directional and the PT takes place on a longer timescale compared with water.[23] Ionized ammonia clusters have been investigated quite extensively and it has been conclusively shown that the ammonium cation solvated with 4 ammonia units in the first solvation shell $(NH_3)_4NH^+_4$ is especially stable.[24–32] Pure ionized methanol clusters were studied as well and the PT reactions following the ionization were identified here.[33–37] The proton transfer was also studied for mixed water-ammonia,[38,39] water-methanol[35] or methanol-ammonia systems.[40]

The ionization induced reactions in methanol/ammonia systems are interesting from a perspective of astrochemistry. Methanol, ammonia and water molecules are all primary hydrogen bonding molecules which have been abundantly detected in interstellar clouds, ices, comets and on the surface of dust grains.[41–45] Their ices have been also suggested to exist on satellites of gas giants[46] and icy bodies in the outer parts of the solar system (see Ref.[47] and references therein). These environments are continuously exposed to highly energetic charged particles and photons. Resulting rich photochemistry and ion-induced chemistry stimulates formation of complex species, e.g. sugars, amino acids, aldehydes or amphiphilic molecules.[48–51] In fact, laboratory experiments simulating the UV processed ice mixtures showed that the presence of ammonia in these ices is essential for a stabilization of light aldehydes in a harsh space environment.[48] Similarly, the importance of methanol presence in astrochemical ices is critical to the production of amphiphilic molecules.[49] In UV irradiated $CH_3OH:NH_3$ 1:1 ices, high concentration of $H_2CO$ and $CH_2OH$ radicals were detected which indicates the activation of C-H bond upon hard UV ionization.[52,53]

What happens upon the ionization of large ammonia clusters doped with a few methanol molecules? We could anticipate that proton transfer will still take place,[40] yet it is unclear in which direction. Will the transfer proceed from ammonia to methanol, the opposite way, or even within the same respective components? Will there be any structural preference in the ground or ionized state? There are also two possible sources of protons in methanol molecules - the OH or CH$_3$ group, leaving either CH$_3$O or CH$_2$OH radicals behind. The transfer of the proton from the CH$_3$ group was predicted to be accompanied by a large barrier[40] yet the process was observed upon the ionization of pure methanol clusters. To answer the above questions, we combine mass spectrometry of molecular clusters generated in a supersonic beam with ab initio calculations. This combination allows us to provide a consistent picture of the reactions and processes induced by the electron impact ionization.

The paper is organized as follows. We first briefly describe the experimental and theoretical methods. Then, we present the experimental results on mass spectrometry of the clusters. The results are interpreted with ab initio calculations both in the ground and ionized states and possible reaction pathways are discussed. Finally, possible reaction mechanisms are discerned by mass spectrometry with isotopically substituted methanols.

# Experimental and Theoretical methods

## Experiment

This study was performed on the cluster beam (CLUB) apparatus in Prague which is a versatile setup allowing for different cluster experiments.[54] The (NH$_3$)$_N$ clusters were prepared by a supersonic expansion of ammonia (Sigma-Aldrich ≥ 99.98%) at the stagnation pressure of $P_0$ = 3 bar through a conical nozzle (55 $\mu$m diameter, 30 deg opening angle, and

2 mm lengths) into the vacuum of ~10$^{-4}$ mbar. The nozzle was kept at the temperature of $T_0$ = 313 K. The ammonia cluster generation and size distributions under various expansion conditions were investigated previously and the modified Hagena's formula was applied to derive the mean cluster size.[55] According to this formula, the mean (NH$_3$)$_N$ cluster size of $\bar{N}$ ≈ 230 was calculated for our present experimental conditions. The cluster size distribution has a log-normal character of the width approximately corresponding to its mean size.

About 2 cm downstream from the nozzle, the clusters passed through a skimmer and entered a 17 cm long chamber which could be filled with the methanol gas (Sigma-Aldrich ≥ 99.8%) at a well-controlled pressure $p(CH_3OH)$ for the pickup experiment. It ought to be mentioned that the methanol displayed pressure was divided by the correction factor of 1.8 as outlined in the ionization pressure gauge manual (571 Ionization Gauge Tube Varian) and also confirmed by independent calibration within our previous experiments.[56,57] Pickup of methanol and other molecules on different clusters was investigated in detail in our previous experiments.[56–58] Methanol (P.A. Lachner), $CD_3OH$ (99.8% Aldrich) and $CH_3OD$ (99.5% Aldrich) were used in the pickup chamber. For the deuterated molecules the system was filled with the species for long time period (hours-days) to passivate the surfaces and avoid H/D exchange. The experiments repeated over a period of several months showed good reproducibility. Also the evaluation procedure discussed below (comparison of mass peaks normalized to $(NH_3)_nH^+$ intensity) would bypass any ambiguity which could arise from the H/D exchange.

The cluster beam then passed through another two differentially pumped vacuum chambers and after about 1.5 m flight path from the skimmer, it entered the ionization region of a perpendicularly mounted reflectron time-of-flight mass spectrometer (TOFMS). The mass spectra were recorded after ionization with an electron beam in a crossed beam arrangement. The mass spectrometer with electron ionization was first implemented and described in detail elsewhere.[59,60] Briefly, the clusters were ionized with 70 eV electrons from an electron gun with a frequency of 5 kHz. The 2 $\mu$s long ionization pulse was followed by 0.5 $\mu$s delay before the ions were extracted by 3 kV pulse into the time-of-flight region, and subsequently they were accelerated to 6 keV. After about 95 cm flight path, the ions were detected with a multichannel plate and the mass spectra were recorded.

## Calculations

All neutral and ionized geometries were optimized at the MP2/6-31++G** level of theory in the Gaussian 09 package[61] and a frequency analysis for the optimized geometries was always performed in order to confirm the energy minima. Subsequently, dissociation energies ($D_E$) for methanol-ammonia clusters were calculated using explicitly correlated CCSD(T)F12a/aug-cc-pVTZ method (or its unrestricted UCCSD(T)-F12a version for the ionized clusters with open-shell electronic structure) and MP2-F12/aug-cc-pVTZ for the protonated ammonia clusters.[62,63] All the energy calculations were performed in the MOLPRO 2015 package.[64,65] The interaction energies were counterpoise corrected (CP) in a monomer rigid scheme, i.e. we neglected the change in the monomer geometry during the complex formation. The CP-corrected energetics with a triple-$\zeta$ basis set and explicitly correlated methods have been shown to provide high-quality energies close to the basis set limit.[66]

Ionization energies were calculated for the optimized neutral geometries at the EOM-IP-CCSD(dT)/aug-cc-pVTZ level in the Q-CHEM 4.3 package.[67,68]

## Results and discussion

### Mass spectra of pure and methanol doped $(NH_3)_N$ clusters

First, we have measured the mass spectra of pure ammonia clusters at the 70 eV electron energy. The spectra are strongly dominated by the protonated $(NH_3)_nH^+$ fragments as illustrated in Fig. 1. The top panel (a) shows relative abundances of $(NH_3)_nH^+$ and $(NH_3)^+_n$ fragments as a function of $n$, and panel (b) below shows a detail of the mass spectrum around $n$ = 10 – 13.

The $(NH_3)_nH^+$ series exhibits a strong magic peak at $n = 5$, explained by filling of the first solvation shell around the protonated ammonia $NH^+_4$.[28–30] Besides, the spectra also contain fragments corresponding to additional hydrogen loss (or gain), and to metastable fragmentation (at fractional masses). All these features have been observed and analyzed in detail previously and the pure ammonia cluster ionization has been understood.[28–32]

It is worth mentioning that in our mass spectrum prevail the $(NH_3)_nH^+$ fragments only up to $n \approx 35$ while the mean neutral cluster size corresponds to $\bar{N} \approx 230$.[55] Some larger fragments were present in the spectrum too, however, with more than ten times smaller intensities than the maximum. This suggests a strong cluster fragmentation upon an electron ionization. Similar effect was observed recently also for water clusters in two different experiments.[69,70] These observations still need a theoretical justification, however, we can exclude possible experimental artifacts in the present experiment. In principle, our TOFMS exhibits some mass discrimination due to the perpendicular arrangement to the cluster beam. However, it has been demonstrated that a mass range of about $10^3$ amu can be covered without discrimination and it can be tuned to large masses of several thousands of mass units. This has been discussed in detail in our previous publication.[69] In the present experiments, when tuning the TOFMS for larger $(NH_3)_nH^+$ fragments, their intensity increased only slightly, yet the intense part of the spectrum still ended at about $n \approx 35$ (see figure S1 in Supporting Information). Another experimental reason for the small fragments could be multiple charging and Coulomb explosion of the clusters.[71] However, the mass spectra measured at low electron energies between 8-15 eV, where only single ionization is possible, were very similar in their overall shape suggesting that multiple charging with 70 eV electrons could not have an overwhelming effect. Although the cluster fragmentation upon an ionization represents an interesting issue and can be further investigated, it does not have any effect on the major conclusions of the present study which concern the uptake of methanol molecules on the $(NH_3)_N$ clusters and the ionization of the mixed $(NH_3)_N(CH_3OH)_M$ clusters.

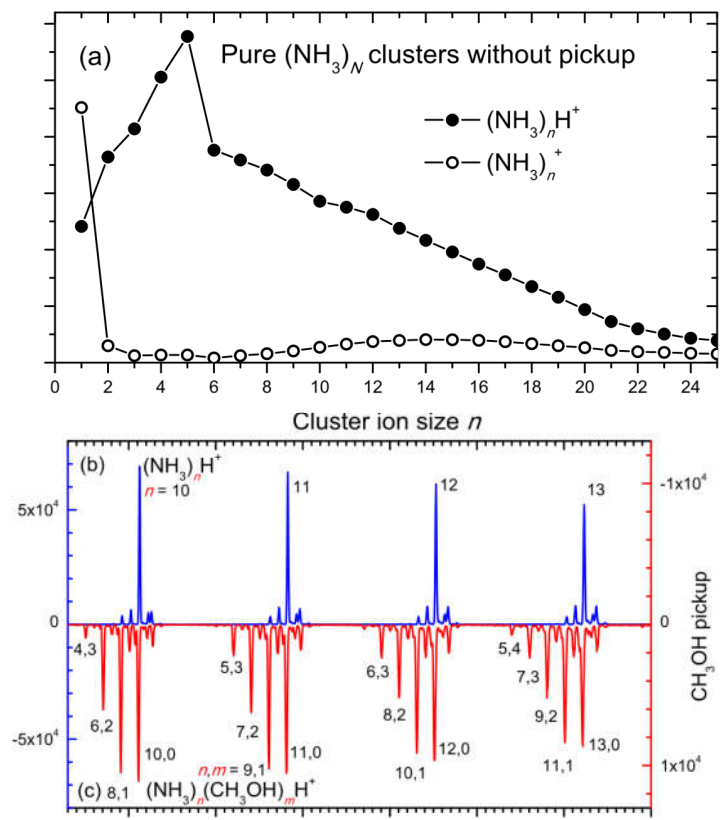

Figure 1: Relative abundances of $(NH_3)_nH^+$ and $(NH_3)^+_n$ fragments in the mass spectrum after 70 eV electron ionization of pure $(NH_3)_N$, $\bar{N} \approx 230$, clusters (a). Detail of the $(NH_3)_N$ mass spectrum (b), and the spectrum with methanol at the pickup pressure $p(CH_3OH) \approx 1.4 \times 10^{-4}$ mbar (c).

The bottom panel in Fig. 1 (c) shows the corresponding mass spectrum recorded at the methanol pressure $p(CH_3OH) \approx 1.4 \times 10^{-4}$ mbar in the pickup cell. The spectrum after the pickup exhibits additional $(NH_3)_n(CH_3OH)_mH^+$ series (labeled by $n,m$). We plot the integrated mass peak intensities for these $(NH_3)_n(CH_3OH)_mH^+$ series for different $m$ in Fig. 2. These dependencies are very similar to the above case of the pure ammonia clusters shown in Fig. 1 (a). The series for $m$ = 0,1 and 2 exhibit the magic peak at $n$ = 5. Further series $m$ = 3 and 4 do not exhibit any maximum at $n$ = 5, yet there is a steeper rise of the spectra to this point where the character of the spectrum changes. Similar dependencies were also measured at other methanol pickup pressures between $10^{-4}$ and $10^{-3}$ mbar. The presence of the magic peak at $n$ = 5 in all these dependencies suggests a similar ionization mechanism for all the series.

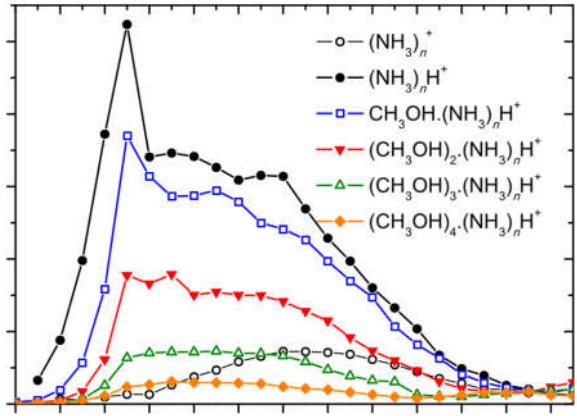

Figure 2: Relative abundances of various $(NH_3)_n(CH_3OH)_mH^+$ series as a function of the number of ammonia molecules $n$. The spectrum recorded after 70 eV electron ionization corresponds to the pickup of methanol on $(NH_3)_N$, $\bar{N} \approx 230$, clusters at the pickup pressure $p(CH_3OH) \approx 1.4 \times 10^{-4}$ mbar.

There are on average 210 ammonia molecules in the clusters and only up to about 10 picked up methanol molecules (as argued below). Therefore, we propose that the ionization of the cluster starts with an $NH_3$ molecule ionization yielding the protonated ammonia $NH_4^+$ most likely solvated with four $NH_3$ molecules. However, the relatively high contribution of the methanol containing $(NH_3)_n(CH_3OH)_mH^+$, $m \geq 1$ mass peaks in the spectra is surprising, considering the high ratio of ammonia molecules with respect to the added methanol. To quantify this ratio for each individual pickup pressure $p(CH_3OH)$, we have integrated the intensities $I_{m,n}$ of $(NH_3)_n(CH_3OH)_mH^+$ series over $n$: $S_m = \sum_n I_{m,n}$ for all observed $m$ series (including $m = 0$). Then these intensities were normalized to the total intensity at a given pickup pressure $p(CH_3OH)$ obtained by summing $S_m$ over $m$: $S = \sum_m S_m$. The normalized intensities $S_m/S$ are plotted in Fig. 3. Clearly, the pure $(NH_3)_nH^+$ series decreases with increasing pickup pressure and the mixed $(NH_3)_n(CH_3OH)_mH^+$ series with larger $m$ contribute more significantly. Qualitatively this is not surprising, however, the relatively high contribution of the mixed $(NH_3)_n(CH_3OH)_mH^+$ series is much higher than expected from the ratio of ammonia and methanol molecules in the clusters.

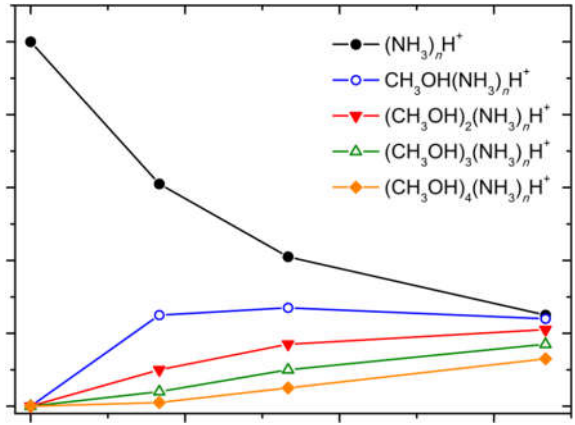

Figure 3: Normalized intensities of different $(NH_3)_n(CH_3OH)_mH^+$ fragments integrated over $n$ at given pickup pressure as a function of $p(CH_3OH)$.

We can roughly estimate the number of adsorbed molecules on the neutral clusters. The mean cluster size under our experimental conditions is $\bar{N} = 230$ $NH_3$ molecules. Assuming the solid ammonia density $\rho = 817$ kg·m³ and close packing of the molecules in the sperical cluster, we can estimate the mean cluster radius $R_N \approx 12.4$ Å. Adding the methanol molecule diameter $R_M = 1.4$ Å, we can estimate the mean geometrical cross section $\sigma \approx 600$ Å$^2$ for the cluster-molecule collisions in the pickup cell. The average number of collisions can be calculated as $n_{col} = L \cdot \sigma \cdot \frac{p}{k_B \cdot T}$, where $L$ = 17 cm is the pickup chamber lengths. At the maximum applied pickup pressure 3.3 × 10$^{-4}$ mbar, this corresponds to $n_{col} \approx 8$. This is in good agreement with the (NH$_3$)$_n$(CH$_3$OH)$_m$H$^+$ series, for which the maximum $m$ = 8 was observed. Based on the calculated binding energies, we can estimate that about two ammonia can evaporate upon a single methanol molecule pickup attachment. Thus, we can assume that the clusters with the most of methanol molecules are composed of about 8 CH$_3$OH and 210 NH$_3$ molecules, i.e. the number of methanol molecules corresponds to less than 4% of the number of ammonia molecules in the cluster. This is just a rough estimate and further factors should be considered. For example, we can assume only the number of ammonia molecules on the cluster surface. However, this is still about 150 out of 210 molecules. On the other hand, the pickup cross section might be larger than the above calculated geometrical cross section of the cluster, as we have shown in the case of water clusters.[56,57,72] For a proper treatment of the pickup processes, Poisson statistics should be considered and convoluted with the size distribution of the neutral clusters. However, such elaborate treatment for the present system seems not justified in the view of large uncertainties of other factors such as the sticking probability of methanol on ammonia clusters and coagulation and fragmentation pattern. Therefore, the present rough estimate based just on the mean cluster size and the mean number of colliding molecules was used to make the major point: the number of methanol molecules in the mixed clusters is only a few percent compared to the number of ammonia molecules.

However, summing up all the contribution of the methanol containing cluster fragments at the highest pickup pressure $p(CH_3OH) = 3.3 \times 10^{-4}$ mbar in Fig. 3, it is 3-times larger than the total contribution of the pure protonated ammonia. Thus 75% fragments contain some methanol, while there is only a few percent of methanol molecules in the clusters. This suggests either a preferential ionization of methanol in the clusters, or a strong propensity of methanol sticking to the ion core of the cluster after the ionization. We show below that the latter explanation is more probable.

## Ab initio perspective

The observed experimental data are interpreted with the ab initio calculations. To get an insight into the character of the formed clusters, we start with a discussion of the structure and energetics of pure and mixed clusters with ammonia and methanol. We exploit dimers and trimers as simple models, allowing us to understand the arrangement in larger systems explored in the experiment. Next, we investigate possible processes following the ionization of the clusters.

The question we address here is whether there is any preferential motif in the ground state of the mixed clusters, i.e. whether we can expect a formation of clusters with a specific composition or arrangement, explaining the reactivity upon the ionization. We therefore calculate binding energies for an ammonia dimer, methanol dimer and mixed dimers between these molecules. The ammonia dimer has been previously found to be a weakly bound, floppy system with two close energy minima[23,73] while the methanol dimer has only a single global energy minimum.[74] The hydrogen bond (HB) in ammonia dimer is somewhat bent with $\angle NHN = 162°$ and the distance of 2.25 Å (Fig. 4a). The binding energy of the complex is only 0.14 eV. The hydrogen bond in the methanol dimer (Fig. 4b) is much stronger, with a binding energy of 0.26 eV. The OH···O arrangement is almost linear ($\angle OHO = 172°$) and the bond is shorter (1.89 Å) than in the ammonia dimer, making it an almost ideal HB. These HB properties propagate into mixed dimers enabling to form two possible configurations with the ammonia or methanol molecule acting as the hydrogen bond donors, AM (Fig. 4c) or MA (Fig. 4d), respectively. The MA conformer appears to be energetically favored as the hydrogen bond is shorter and twice as much energy is needed to dissociate it. It therefore follows from the dimer model that methanol tends to cluster with other methanol molecules or acts as hydrogen bond donor for ammonia molecules. Similar binding characteristics are found for trimers (see the Supporting Information). In the larger clusters, generally we find it easier to dissociate ammonia clusters rather than the methanol clusters. Fig. 4 also shows the first few ionization energies and the ionized moiety is marked. More detailed analysis is provided in the Supporting Information.

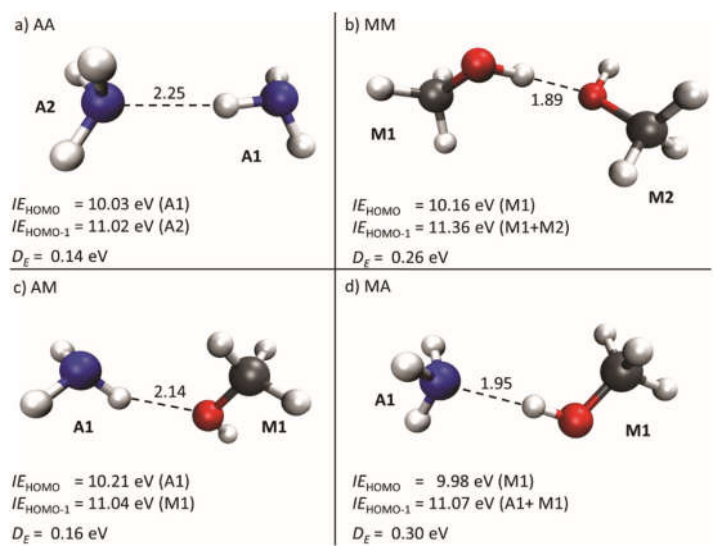

Figure 4: Structures and energetics of the neutral ground state of: a) ammonia dimer, b) methanol dimer, c) ammonia-methanol dimer and d) methanol-ammonia dimer. Ionization energies were calculated at EOM-IP-CCSD(dT)/aug-cc-pVTZ level and dissociation energies ($D_E$) at CCSD(T)-F12a/aug-cc-pVTZ level with the counterpoise correction. The hydrogen bonds distances are in Å units and A1, A2, M1 and M2 denotes individual ammonia (A) or methanol (M) molecules.

Next, we investigate the possible clusters formed upon the ionization. We focus on the mixed ammonia-methanol dimers (Fig. 5) as the proton transfer reactions in pure clusters are already well documented. We identified three possible structures of ionized ammoniamethanol dimers. In all cases, ammonia or ammonium cation were acting as the hydrogen bond donor and the charge always ends up on the ammonia moiety. In the first two cases, the proton is transferred from the methanol molecule to ammonia; the proton can originate from both the hydroxyl (Fig. 5a) or methyl group (Fig. 5b) of the methanol molecule. The resulting structure has the positive charge located on a NH$^+_4$ moiety and an unpaired electron is either on the oxygen or carbon atom. In fact, the structure with the proton transferred from the methyl group appears to be the global minimum. Similarly, the proton transferred from the methyl group has been found in the ionized methanol clusters already by Lee et al.[75] The third geometry (Fig. 5c) with nearly planar NH$_3$ radical cation is energetically the least stable. This structure has a large dissociation energy and very short intermolecular bond of 1.43 Å, i.e. the two units essentially merged into a single molecule.

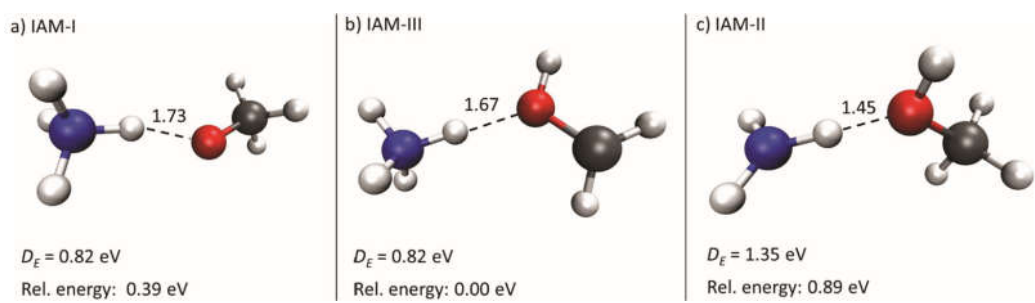

Figure 5: Structures and energetics of the ground ionized state of methanol-ammonia dimer. Dissociation energies (D$_E$) were calculated at CCSD(T)-F12a/aug-cc-pVTZ level with the counterpoise correction. The hydrogen bonds distances are in Å units and A1, A2, M1 and M2 denotes individual ammonia (A) or methanol (M) molecules.

Similar conclusions can be drawn from the investigation of the ionized trimer clusters. In the ionized methanol-ammonia-ammonia (IMAA) and ionized methanol-methanol-ammonia (IMMA) trimers, all the optimized structures are branched with the linking molecule being either the NH$^+_4$ cation (Fig.6a, b, c and Fig.7a, b, c), a methanol molecule (Fig.6e and Fig. 7d), or CH$_2$OH (Fig. 6d and Fig. 7e) and CH$_3$O radical (Fig. 6f). The proton can be transferred again from a hydroxyl or methyl group. The two lowest energy minima for both systems have the proton transferred from methyl groups. The energy difference between the two most stable structures is below 0.1 eV. The largest dissociation energies were found when the NH$^+_4$ cation removed from the cluster (Fig. 6d,e and Fig. 7d) while the lowest dissociation energy is attributed to the NH$_3$ molecule facing the methyl group (Fig. 6f). This atypical bonding is also the highest lying energy-minimum geometry. Structures with a protonated methanol were not found in any of the ionized geometries.

Based on the energetics of the small clusters, we might assume that the charge will always be localized on the ammonia units, forming $NH_4^+$ cation solvated either with methanol or other ammonia molecules. Our calculations predict that the proton can originate either from ammonia molecules (in that case $NH_2$ radical will be boiled off the cluster), from the O-H group of the methanol molecule (with $CH_3O$ radical leaving the cluster) or from the $CH_3$ group of methanol (and $CH_2OH$ leaving the cluster). The two processes with a proton transfer along the O-H or N-H bond should be essentially ballistic, in analogy with water and ammonia clusters.[23,76] On the other hand, the transfer from the $CH_3$ group should be accompanied by an energy barrier.[40] In the next section, we experimentally address the question whether the proton transfer takes place also form the methyl group.

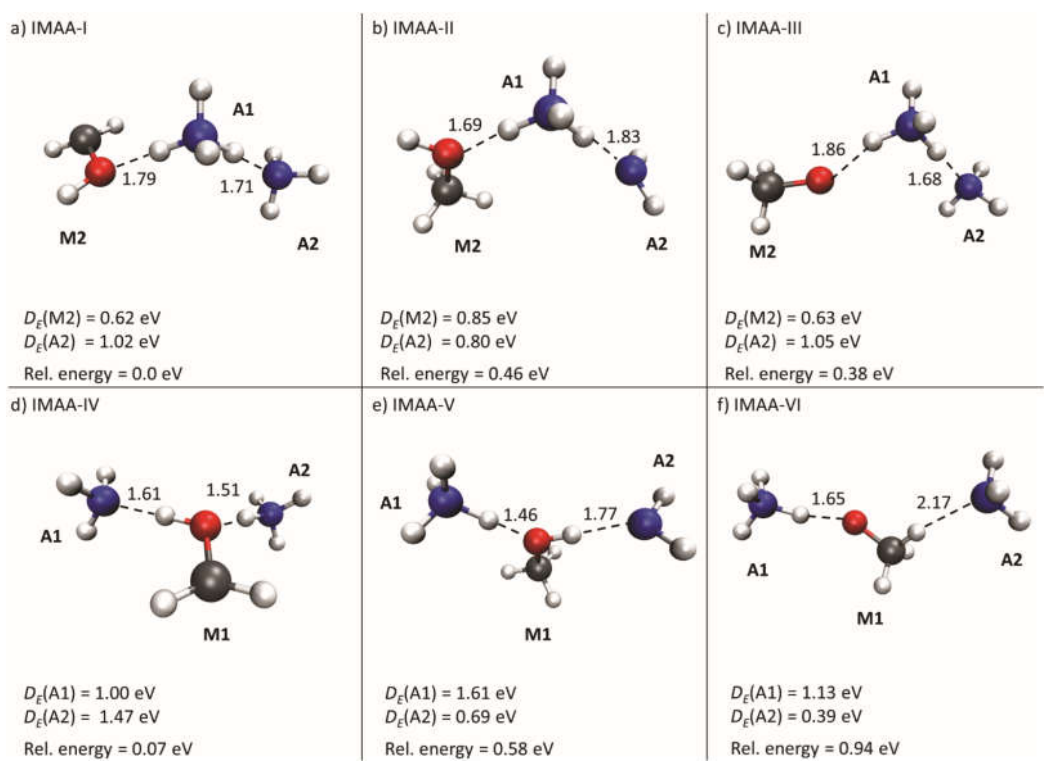

Figure 6: Structures and energetics of the ground ionized state of methanol-ammoniaammonia trimer. Dissociation energies ($D_E$) were calculated at CCSD(T)-F12a/aug-ccpVTZ level with the counterpoise correction. The hydrogen bonds distances are in Å units and A1, A2, M1 and M2 denotes individual ammonia (A) or methanol (M) molecules.

In all the experiments, we have observed the "magic number" cluster ion $(NH_3)_5H^+$. It is reasonable to assume that the ammonium cation with its first solvation layer comprising of 4 ammonia units is almost always present in the system and the remaining molecules are less bound. This conjecture is supported by the ab initio calculations. We have already seen that the proton will most likely end up on the ammonia unit in a form of an ammonium cation. Further insight can be gained from Table 1. We observed that the ammonia molecule in the $(NH_3)_5NH^+_4$ cluster is significantly less bound than the first 4 ammonia units (reactions (a) and (b), respectively). This is in agreement with previous measurements: the calculated values agree well with the reaction enthalpies estimated by Searles and Kebarle.[77] They are also reasonably close to the binding energies given in a more recent reference.[78]

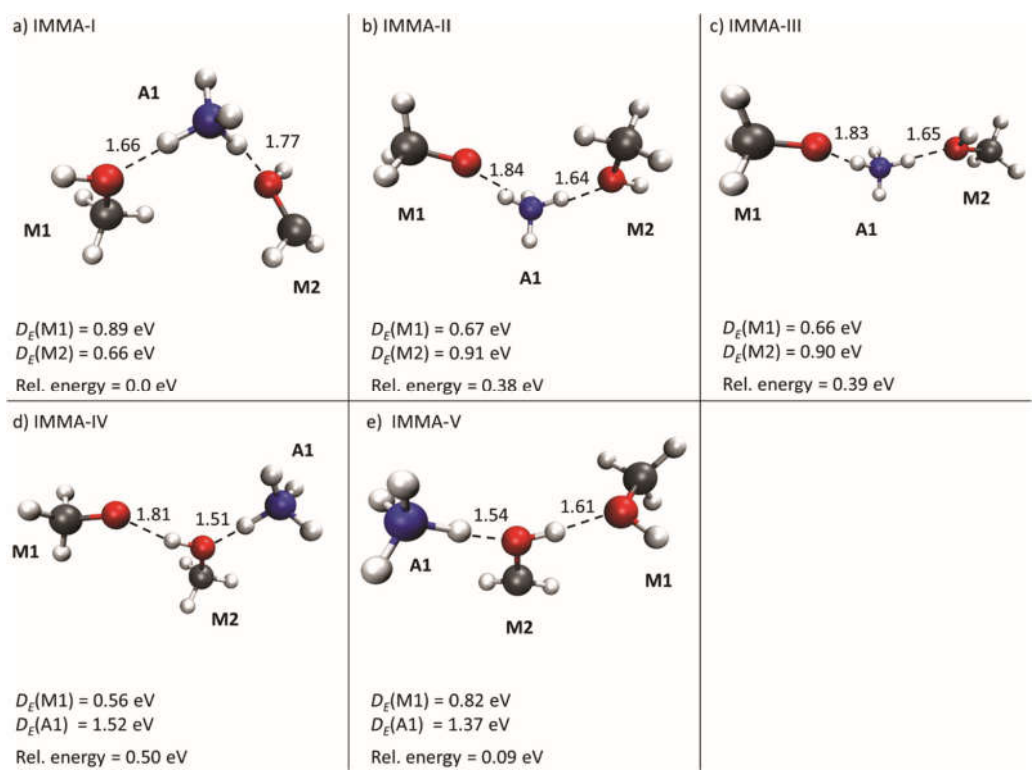

Figure 7: Structures and energetics of the ground ionized state of methanol-ammoniaammonia trimer. Dissociation energies ($D_E$) were calculated at CCSD(T)-F12a/aug-ccpVTZ level with the counterpoise correction. The hydrogen bonds distances are in Å units and A1, A2, M1 and M2 denotes individual ammonia (A) or methanol (M) molecules.

At the same time, the methanol molecule is more bound in the $CH_3OH(NH_3)_4NH^+_4$ cluster than ammonia unit is in the $(NH_3)_5NH^+_4$ cluster (reaction (b) and (c), respectively). We also observe that from he energy perspective, the ammonia in the first solvation shell will not be replaced by the methanol molecule (reaction (d)). We can therefore conclude that whatever molecule is ionized, we end up with solvated ammonium cation which then preferentially binds methanol molecules.

Table 1: Dissociation energies of ammonia and methanol units from different protonated ammonia-methanol clusters with $NH^+_4$ ion. Calculated at the MP2F12/aug-cc-pvtz level with the BSSE correction. Square brackets denote first solvation sphere around the $NH^+_4$ cation. Corresponding geometries can be found in the supporting information.

| Reaction | Energy / eV |
|---|---|
| a) $[NH_4(NH_3)_4]^+ \rightarrow [NH_4(NH_3)_3]^+ + NH_3$ | 0.61 |
| b) $[NH_4(NH_3)_4]^+NH_3 \rightarrow [NH_4(NH_3)_4]^+ + NH_3$ | 0.35 |
| c) $[NH_4(NH_3)_4]^+CH_3OH \rightarrow [NH_4(NH_3)_4]^+ + CH_3OH$ | 0.48 |
| d) $[NH_4(NH_3)_4]^+CH_3OH \rightarrow [NH_4(NH_3)_3CH_3OH]^+NH_3$ | 0.12 |

## Tracing proton transfer paths experimentally

The theoretical calculations above suggests that the ionization of methanol in the clusters can be followed by a proton transfer from methanol to form $NH^+_4$. An interesting possibility, which in theory has the largest exorgicity, is the proton transfer from the methyl group. To test experimentally whether this process occurs, we can use partially deuterated methanol $CD_3OH$. The proton transfer from the methyl group of $CD_3OH$ would yield the deuterated fragments $(NH_3)_nD^+$ and $(NH_3)_n(CD_3OH)_mD^+$ in the spectra. On the other hand, if the proton transfer occurred from the hydroxyl group, we could use $CH_3OD$ and observe the

$(NH_3)_nD^+$ and $(NH_3)_n(CH_3OD)_mD^+$ fragments.

Fig. 8 shows the comparison between the mass spectra with the pickup of $CD_3OH$ (top blue spectrum) and $CH_3OH$ (bottom red spectrum). The series $(NH_3)_n(CD_3OH)_mH^+$ in the top spectrum (labeled blue) would correspond to the $(NH_3)_n(CH_3OH)_mH^+$ series in the bottom spectrum (labeled red), assuming that the proton is transferred from ammonia or from the OH group of methanol. However, in the top spectrum, this $(NH_3)_n(CD_3OH)_mH^+$ series would also coincide with the $(NH_3)_{n+2}(CD_3OH)_{m-1}D^+$ fragments, i.e. the series which would result from the deuterium transfer from the $CD_3$ group of methanol. In particular, the $(NH_3)_n(CD_3OH)H^+$ series would coincide with the $(NH_3)_{n+2}D^+$ fragments.

To determine a possible contribution of the $D^+$ transfer from the $CD_3$ group of methanol,

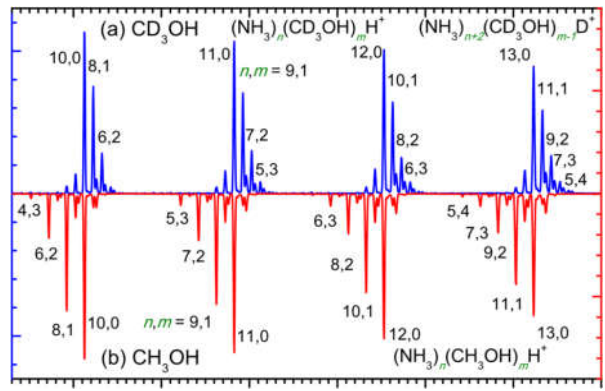

Figure 8: Comparison of the mass spectra after the $(NH_3)_N$, $N$- ≈ 230 of deuterated methanol $CD_3OH$, top (a), and normal methanol $CH_3OH$, bottom (b), at the same pickup pressure $0.8 \times 10^{-4}$ mbar.

we need to disentangle the contribution of the $(NH_3)_{n+2}D^+$ fragments. Fig. 9 shows the integrated intensities of the $(NH_3)_n(CH_3OH)_mH^+$ and the equivalent $(NH_3)_n(CD_3OH)_mH^+$ series showing the corresponding mass peaks with the same $n,m$ next to each other. The x-axis indicates $n$ for the $(NH_3)_nH^+$ fragments. The red bars correspond to the $CH_3OH$ pickup spectrum in Fig 8 (b) (bottom red). Each group of peaks has been normalized to the corresponding $(NH_3)_nH^+$ fragment intensity. The blue bars correspond to the intensities of the mass peak series in the $CD_3OH$ pickup spectrum in Fig. 8 (a) (top blue). The intensities have also been normalized on $(NH_3)_nH^+$ fragment intensities in this spectrum. We assume that the $(NH_3)_nH^+$ intensities in both spectra would be the same under identical conditions. In other words, we assume that the $(NH_3)_n(CD_3OH)_mH^+$ series intensities in the top spectrum follow the $(NH_3)_n(CH_3OH)_mH^+$ intensities of the equivalent series in the bottom spectrum. This is a justified assumption unless the ionization in the deuterated system would run via completely different pathways, which is unlikely. Thus, subtracting the normalized intensities (red from blue), would yield the contribution from the $(NH_3)_n(CD_3OH)_mD^+$ series corresponding to the $D^+$ transfer process from the $CD_3$ group. These differences are indicated by the green bars. Clearly, they are negligible within our experimental error (of less than 15%). Thus there is no clear indication for the $(NH_3)_nD^+$ and $(NH_3)_n(CD_3OH)_mD^+$ fragments. The experiments were also performed at other pickup pressures with the same results. In summary, there is no indication for the $D^+$ transfer from the methyl group of methanol.

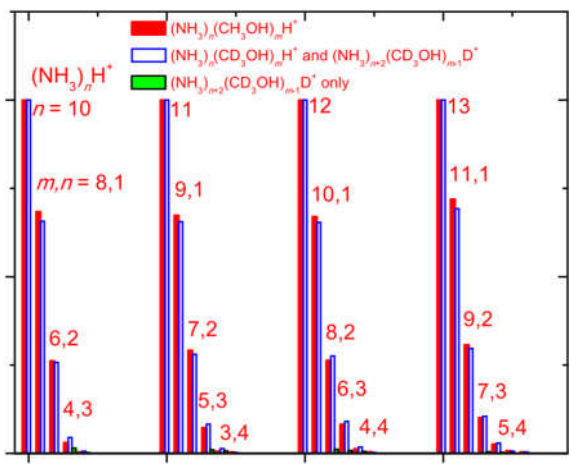

Figure 9: Comparison of the mass peak intensities of $(NH_3)_n(CH_3OH)_mH^+$ (red) and the equivalent $(NH_3)_n(CD_3OH)_mH^+$ (blue) series as a function of $n,m$. The difference (green) corresponds to the $(NH_3)_{n+2}(CD_3OH)_{m-1}D^+$ fragments -their intensities are negligible within the experimental error.

Analogous experiments were performed with the pickup of $CH_3OD$ to search for the $D^+$ transfer from the hydroxyl group of methanol. The results were essentially the same as in the case of $CD_3OH$, however, their interpretation is hampered for $CH_3OD$ by the mass coincidence of $(NH_3)_nH^+$ fragments with $(NH_3)_{n-2}(CH_3OD)D^+$ ions. Due to this coincidence, the normalization of the mass peak intensities on $(NH_3)_nH^+$ is not unambiguous. Therefore we present the data and corresponding analysis only in the Supporting Information where it is shown that following the above procedure of data analysis would yield again a very small contribution of $D^+$ transfer at the experimental error limit. However, due to the above mentioned ambiguity of the procedure for the $CH_3OD$ case, we cannot rule out some contribution of the $D^+$ transfer from the hydroxyl group based on this analysis.

The $D^+$-transfer in the mixed $CH_3OD/NH_3$ clusters was observed previously.[79] However, it should be noted that quite different experiment and clusters were investigated in that study. The mixed $CH_3OD/NH_3$ species were produced in coexpansion where relatively small clusters were generated (compared to the present pickup study). Besides, in the mixed expansions the pure ammonia and pure methanol as well as the mixed clusters could be generated and contribute to the mass spectra. On the other hand, in the present study much larger ammonia clusters are generated and can pick up some methanol molecules only after their structure has been stabilized by evaporation of ammonia molecules in the collision free region ("zone of silence") of the expansion. Also the ionization method was different: multiphoton ionization at 355 nm, and there is quite some evidence in the literature that

electron ionization and photoionization can lead to quite different intracluster reactions.[80–82]

The mixed $CH_3OD/NH_3$ clusters generated in a coexpansion were also investigated in a pump-probe experiment.[83] Yamada et al. detected the small protonated cluster ions $NH^+_4 (CH_3OH)_m(NH_3)_n$, measured their time evolution and analyzed it in terms of a kinetic model consisting of three-step dynamics: generation of a radical pair $(NH_4–NH_2)^*$, followed by the relaxation process of radical-pair clusters, and subsequent dissociation of the solvated $NH_4$ clusters. They have not observed any significant hydrogen transfer between ammonia and methanol. However, it ought to be mentioned that the ion chemistry in our present experiments may differ from the neutral radical chemistry discussed by Yamada et al.[83]

## Conclusions

Several methanol molecules were picked up on large ammonia clusters $(NH_3)_N$ with a mean size of $\bar{N} \approx 230$. Subsequent electron ionization of the mixed $(NH_3)_N(CH_3OH)_M$ clusters resulted in the $(NH_3)_n(CH_3OH)_mH^+$ ions where the contribution of methanol containing ion fragments was unexpectedly high compared to the pure protonated ammonia $(NH_3)_nH^+$. At the highest exploited pickup pressures, the methanol containing ion fragments $(NH_3)_n(CH_3OH)_mH^+$ contributed approximately 75% to the integrated ion yield compared to 25% contribution of $(NH_3)_nH^+$ ions, while the average composition of the neutral $(NH_3)_N(CH_3OH)_M$ clusters was estimated to be roughly $N:M \approx 210:10$.

Based on this clear disproportion between the neutral and ionized species composition, we proposed an ionization mechanism in which preferentially the ammonia is ionized in the first place resulting in the $NH^+_4$ ion core (most likely solvated with 4 ammonia molecules yielding the well-known "magic number" $(NH_3)_4NH^+_4$). The methanol molecules then show strong propensity for sticking to the fragment ion resulting in the mixed $(NH_3)_n(CH_3OH)_mH^+$ ions. Even if methanol were ionized, the same product ions would be generated. Our theoretical calculations performed for small mixed clusters (dimers and trimers of ammonia and methanol) supported these observations. They demonstrated significantly stronger binding energies of methanol compared to ammonia in both neutral and ionized clusters.

The proposed mechanism is in agreement with the data found previously for the analogical water-ammonia systems.[84,85] In the $(NH_3)_n(H_2O)_mH^+$ clusters the proton was found to be solvated within the first solvation shell by ammonia yet the small protonated clusters prefer binding with water molecules beyond the first solvation shell. Such a mechanism was found both in the molecular beam mass spectrometry studies[85] as well as in the previous high pressure mas spectrometry experiments.[84] The binding energies of methanol and water are comparable[78] and it is therefore reasonable to expect analogical mechanism for the present system.

We have also investigated possible intracluster reactions with a proton being transferred from methanol. The ab initio calculations suggested an energetically favourable ionization mechanism in which a proton is transferred from the methyl group of methanol. To test this mechanism, experiments with partially deuterated methanols were performed. These experiments proved no significant contribution of the proton transfer from the methyl group of methanol. Due to the mass coincidences, the experiments with $CH_3OD$ could not be interpreted unambiguously. Therefore, we could not exclude completely nor confirm unambiguously some contribution of a proton transfer from the hydroxyl group.

Also worth noting, is the observed strong fragmentation of the clusters upon the electron ionization. This is an interesting experimental fact which calls for further investigation and theoretical justification.

The observations in this work may be relevant for ionization of ammonia ices with adsorbed molecules in the space. The probability of generation of mixed ions in such processes might be much higher than what one would predict simply based on a low coverage of ammonia ices by foreign molecules.

## Acknowledgement

The authors thank to the support of the Czech Science Foundation grant. No.: 17-04068S and the Praemium Academiae 2018 of the Czech Academy of Sciences, the Austrian Science Fund (FWF) project No.: W1259-N27 "DK-ALM". Also fruitfull discussions and ideas from J. Fedor are acknowledged.

## Supporting Information Available

Further experimental details, the experimental results for the pickup of methanol with deuterated hydroxyl group $CH_3OD$ on ammonia clusters $(NH_3)_N$, the Cartesian coordinates for all the optimized structures, computational results for protonated ammonia and larger neutral ammonia/methanol clusters are shown in the Supporting Information.

This material is available free of charge via the Internet at http://pubs.acs.org/.

## Graphical TOC Entry

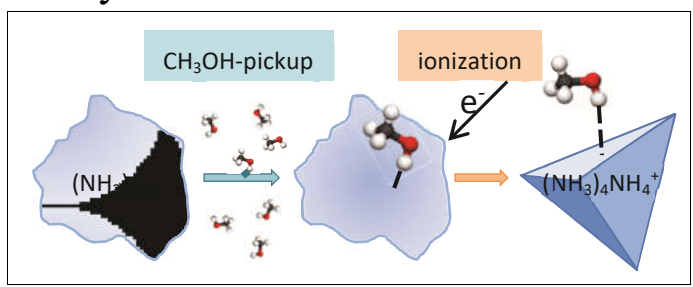